\documentclass{article}
\usepackage{graphicx}
\usepackage{amsmath}
\usepackage{amsfonts}
\usepackage{amssymb}

\begin{document}

\author{Antony Valentini\\Augustus College}

\begin{center}
{\LARGE Subquantum Information and Computation}\footnote{To appear in:
\textit{Proceedings of the Second Winter Institute on Foundations of Quantum
Theory and Quantum Optics: Quantum Information Processing}, ed. R. Ghosh
(Indian Academy of Science, Bangalore, 2002).}

\bigskip

\bigskip\bigskip

\bigskip Antony Valentini\footnote{email: a.valentini@ic.ac.uk}

\bigskip

\bigskip

\textit{Theoretical Physics Group, Blackett Laboratory, Imperial College,
Prince Consort Road, London SW7 2BZ, England.\footnote{Corresponding address.}}

\textit{Center for Gravitational Physics and Geometry, Department of Physics,
The Pennsylvania State University, University Park, PA 16802, USA.}

\textit{Augustus College, 14 Augustus Road, London SW19 6LN,
England.\footnote{Permanent address.}}

\bigskip
\end{center}

\bigskip

\bigskip

\bigskip

It is argued that immense physical resources -- for nonlocal communication,
espionage, and exponentially-fast computation -- are hidden from us by quantum
noise, and that this noise is not fundamental but merely a property of an
equilibrium state in which the universe happens to be at the present time. It
is suggested that `non-quantum' or nonequilibrium matter might exist today in
the form of relic particles from the early universe. We describe how such
matter could be detected and put to practical use. Nonequilibrium matter could
be used to send instantaneous signals, to violate the uncertainty principle,
to distinguish non-orthogonal quantum states without disturbing them, to
eavesdrop on quantum key distribution, and to outpace quantum computation
(solving $NP$-complete problems in polynomial time).

\bigskip

\bigskip

\bigskip

\bigskip

\bigskip

\bigskip

\bigskip

\bigskip

\bigskip

\bigskip

\bigskip

\bigskip

\bigskip

\bigskip

\bigskip

\bigskip

\bigskip

\bigskip

\bigskip

\bigskip

\bigskip

\bigskip

\bigskip

\bigskip

\bigskip

\bigskip

\section{\bigskip Introduction and Motivation}

In quantum theory the Born probability rule is regarded as a fundamental law
of Nature: a system with wavefunction $\psi$ has an associated probability
distribution $\rho=|\psi|^{2}$. However, there are reasons to believe that
this distribution is not fundamental, but merely corresponds to a special
`equilibrium' state, analogous to thermal equilibrium [1--7]. For there seems
to be a `conspiracy' in the known laws of physics: long-distance quantum
correlations suggest that our universe is fundamentally nonlocal, and yet the
nonlocality cannot be used for practical instantaneous signalling. This
apparent conspiracy may be explained if one supposes that signal-locality is
merely a property of the special state $\rho=|\psi|^{2}$, in which nonlocality
happens\textit{ }to be hidden by quantum noise; while for a general
distribution $\rho\neq|\psi|^{2}$, nonlocality would be directly visible.
While $\rho=|\psi|^{2}$ to high accuracy now (for all systems probed so far),
perhaps $\rho\neq|\psi|^{2}$ in the early universe, the relaxation
$\rho\rightarrow|\psi|^{2}$ having taken place soon after the big bang. Thus
our experience happens to be restricted to an equilibrium state $\rho
=|\psi|^{2}$ in which locality and uncertainty \textit{appear} to be fundamental.

A heuristic analogy may be drawn with physics in a universe that has reached a
state of thermal `heat death', in which all systems have the same temperature
[2]. In such a universe there is a universal probability distribution given by
the Boltzmann rule $\rho=e^{-E/kT}/Z$, analogous to our universal Born rule
$\rho=|\psi|^{2}$; all systems are subject to a universal thermal noise,
analogous to our universal uncertainty noise; and it is impossible to convert
thermal energy into useful work, just as it is impossible in our universe to
convert quantum nonlocality into a useful instantaneous signal.

A precise model of this scenario is obtained in deterministic hidden-variables
theories such as the pilot-wave theory of de Broglie and Bohm [1--14]. These
nonlocal theories allow one to discuss the properties of hypothetical
nonequilibrium distributions $\rho\neq|\psi|^{2}$, for which it may be shown
that there are instantaneous signals at the statistical level [2, 15, 16].
Thus in these theories it may be asserted that quantum theory is just the
theory of a special state $\rho=|\psi|^{2}$ in which nonlocality happens to be
hidden by statistical noise. And in pilot-wave theory at least, the relaxation
$\rho\rightarrow|\psi|^{2}$ may be accounted for by an \textit{H}-theorem [1,
5], much as in classical statistical mechanics, so that $\rho=|\psi|^{2}$ is
indeed merely an equilibrium state.\footnote{Other authors tend to consider
pilot-wave theory in equilibrium alone. This is like considering classical
mechanics only in thermal equilibrium.}

Here we shall work with the pilot-wave model. The details of that model may or
may not be correct: but it has qualitative features, such as nonlocality, that
are known to be properties of all hidden-variables theories; and it is helpful
to work with a specific, well-defined theory. In this model, a system with
wavefunction $\psi(x,t)$ has a configuration $x(t)$ whose velocity is
determined by $\dot{x}(t)=j(x,t)/|\psi(x,t)|^{2}$, where $j$ is the quantum
probability current. Quantum theory is recovered if one assumes that an
ensemble of systems with wavefunction $\psi_{0}(x)$ begins with a `quantum
equilibrium' distribution of configurations $\rho_{0}(x)=|\psi_{0}(x)|^{2}$ at
$t=0$ (guaranteeing $\rho(x,t)=|\psi(x,t)|^{2}$ for all $t$). In effect, the
Born rule is assumed as an initial condition. But the theory also allows one
to consider arbitrary `nonequilibrium' initial distributions $\rho_{0}%
(x)\neq|\psi_{0}(x)|^{2}$, which violate quantum theory [1--7], and whose
evolution is given by the continuity equation%
\[
\frac{\partial\rho(x,t)}{\partial t}+\nabla\cdot(\dot{x}(t)\rho(x,t))=0
\]
(the same equation that is satisfied by $|\psi(x,t)|^{2}$).

Our working hypothesis, then, is that $\rho=|\psi|^{2}$ is an equilibrium
state, analogous to thermal equilibrium in classical mechanics. This state has
special properties -- in particular locality and uncertainty -- which are not
fundamental. It then becomes clear that a lot of new physics must be hidden
behind quantum noise, physics that is unavailable to us only because we happen
to be trapped in an equilibrium state.

This new physics might be accessible if the universe began in nonequilibrium
$\rho\neq|\psi|^{2}$. First, in theories of cosmological inflation, early
corrections to quantum fluctuations would change the spectrum of primordial
density perturbations imprinted on the cosmic microwave background [6, 7].
Second, relic cosmological particles that decoupled at sufficiently early
times might still be in quantum nonequilibrium today, violating quantum
mechanics [3--7].

The second possibility is particularly relevant here. If relic nonequilibrium
matter from the early universe was discovered, what could we do with it?
Thermal and chemical nonequilibrium have myriad technological applications; we
expect that quantum nonequilibrium would be just as useful.

\section{Detection and Use of Quantum Nonequilibrium}

First we need to consider how a nonequilibrium distribution $\rho\neq
|\psi|^{2}$ might be deduced by statistical analysis of a random sample of
relic matter [7].

Consider the unrealistic but simple example of a large number $N$ of Hydrogen
atoms in the ground state $\psi_{100}(r)$. Assume they make up a cloud of gas
somewhere in space. Because the phase of $\psi_{100}$ has zero gradient, the
de Broglie-Bohm velocity field vanishes, and pilot-wave theory predicts that
each electron is at rest relative to its nucleus. We then have a static
distribution $\rho(r)$, which may or may not equal the quantum equilibrium
distribution $\rho_{eq}(r)=\left|  \psi_{100}(r)\right|  ^{2}\propto
e^{-2r/a_{0}}$. To test this, one could draw a random sample of $N%
\acute{}%
$ atoms ($N%
\acute{}%
<<N$) and measure the electron positions. The sample $r_{1},\;r_{2}%
,\;r_{3},.....,\;r_{N%
\acute{}%
}$ may be used to make statistical inferences about the parent distribution
$\rho(r)$. In particular, one may estimate the likelihood that $\rho
(r)=\rho_{eq}(r)$. Should one deduce that, almost certainly, the cloud as a
whole has a nonequilibrium distribution $\rho(r)\neq\rho_{eq}(r)$, the rest of
the cloud may then be used as a resource for new physics.

For example, one could test $\rho(r)$ via the sample mean $\bar{r}$. If
$\rho(r)$ has mean $\mu$ and variance $\sigma^{2}$, the central limit theorem
tells us that for large $N%
\acute{}%
$ the random variable $\bar{r}$ has an approximately normal distribution with
mean $\mu$ and variance $\sigma^{2}/N%
\acute{}%
$. We can then calculate the probability that $\bar{r}$ differs from $\mu$,
and we can test the hypothesis that $\rho(r)=\rho_{eq}(r)$ with $\mu=\mu
_{eq}=\frac{3}{2}a_{0}$. A standard technique is to compare the probability
$P(\bar{r}|\rho_{eq})$ of obtaining $\bar{r}$ from a distribution $\rho_{eq}$
with the probability $P(\bar{r}|\rho_{noneq})$ of obtaining $\bar{r}$ from
some nonequilibrium distribution $\rho_{noneq}$. One usually refers to
$P(\bar{r}|\rho_{eq})$ and $P(\bar{r}|\rho_{noneq})$ as the `likelihoods' of
$\rho_{eq}$ and $\rho_{noneq}$ respectively, given the sample mean $\bar{r}$.
If $P(\bar{r}|\rho_{eq})<<P(\bar{r}|\rho_{noneq})$, one concludes that
nonequilibrium is much more likely. Similarly, using standard techniques such
as the chi-square test, one may deduce the most likely form of the parent
distribution $\rho(r)$, which almost certainly applies to the rest of the
cloud.\footnote{The same reasoning applies if the parent distribution is
time-dependent: if the sampling is done at time $t_{0}$, and statistical
analysis favours a distribution $\rho(r,t_{0})$ at $t_{0}$, then the most
likely distribution at later times may be calculated by integrating the
continuity equation.}

In what follows, then, we assume that at $t=0$ we have a large number of
particles with the same known wavefunction $\psi_{0}(x)$, and with positions
$x$ that have a \textit{known} nonequilibrium distribution $\rho_{0}%
(x)\neq\left|  \psi_{0}(x)\right|  ^{2}$.

\section{Instantaneous Signalling}

The most obvious application of such `non-quantum' matter would be for
instantaneous signalling across space [7].

Suppose we take pairs of nonequilibrium particles and prepare each pair in an
entangled state $\psi(x_{A},x_{B},t_{0})$ at time $t_{0}$ (by briefly
switching on an interaction). Given the details of the preparation, we may use
the Schr\"{o}dinger equation to calculate the evolution of the wavefunction of
each pair, from $\psi(x_{A},x_{B},0)=\psi_{0}(x_{A})\psi_{0}(x_{B})$ at $t=0$
to $\psi(x_{A},x_{B},t_{0})$ at $t=t_{0}$. We then know the de Broglie-Bohm
velocity field throughout $(0,t_{0})$, and so we may use the continuity
equation to calculate the evolution of the joint distribution for the pairs
from $\rho(x_{A},x_{B},0)=\rho_{0}(x_{A})\rho_{0}(x_{B})$ at $t=0$ to
$\rho(x_{A},x_{B},t_{0})\neq\left|  \psi(x_{A},x_{B},t_{0})\right|  ^{2}$ at
$t=t_{0}$.\footnote{If the velocity field does not vary too rapidly in
configuration space and the time interval $(0,t_{0})$ is not inordinately
long, relaxation to equilibrium will not be significant.} We then have the
situation discussed in detail elsewhere [2]. The marginal distribution
$\rho_{A}(x_{A},t_{0})\equiv\int dx_{B}\,\rho(x_{A},x_{B},t_{0})$ at $A$ is
known, and its subsequent evolution will depend instantaneously on
perturbations applied at $B$, however remote $B$ may be from $A$. Thus
instantaneous signals may be sent from $B$ to $A$.

It might be thought that superluminal signals would necessarily lead to causal
paradoxes. However, it could well be that at the nonlocal hidden-variable
level there is a preferred slicing of spacetime, labelled by a time parameter
that defines a fundamental causal sequence [3, 7, 17,
18].\footnote{Instantaneous signals would define (operationally) an absolute
simultaneity; `backwards-in-time' effects generated by Lorentz transformations
would be fictitious, moving clocks being incorrectly synchronised if one
assumes isotropy of the speed of light in all frames [3, 7, 17].}

\section{Subquantum Measurement}

Let us now consider how our nonequilibrium particles could be used to perform
novel measurements on ordinary, equilibrium systems [7].

Assume once again that we have an ensemble of what we shall now call
`apparatus' particles with known wavefunction $g_{0}(y)$ and known
\textit{nonequilibrium} distribution $\pi_{0}(y)\neq\left|  g_{0}(y)\right|
^{2}$. (The position $y$ may be regarded as a `pointer' position.) And let us
now use them to measure the positions of ordinary `system' particles with
known wavefunction $\psi_{0}(x)$ and known \textit{equilibrium} distribution
$\rho_{0}(x)=\left|  \psi_{0}(x)\right|  ^{2}$. We shall see that, if the
apparatus distribution $\pi_{0}(y)$ were arbitrarily narrow, we could measure
the system position $x_{0}$ without disturbing the system wavefunction
$\psi_{0}(x)$, to arbitrary accuracy, in complete violation of the uncertainty principle.

We shall illustrate the idea with an exactly-solvable model. At $t=0$, we take
a system particle and an apparatus particle and switch on an interaction
between them described by the Hamiltonian $\hat{H}=a\hat{x}\hat{p}_{y}$, where
$a$ is a coupling constant and $p_{y}$ is the momentum canonically conjugate
to $y$. (This is the standard interaction Hamiltonian for an ideal quantum
measurement of $x$ using the pointer $y$.) For simplicity, we neglect the
Hamiltonians of $x$ and $y$ themselves.\footnote{This might be justified by
assuming $a$ to be relatively large; or, one can just accept the above
Hamiltonian as a simple illustrative model.} For $t>0$ the joint wavefunction
$\Psi(x,y,t)$ satisfies the Schr\"{o}dinger equation%
\[
\frac{\partial\Psi(x,y,t)}{\partial t}=-ax\frac{\partial\Psi(x,y,t)}{\partial
y}%
\]
while $\left|  \Psi(x,y,t)\right|  ^{2}$ obeys the continuity equation%
\[
\frac{\partial\left|  \Psi(x,y,t)\right|  ^{2}}{\partial t}+ax\frac
{\partial\left|  \Psi(x,y,t)\right|  ^{2}}{\partial y}=0
\]
The hidden-variable velocity fields $\dot{x}$ and $\dot{y}$ must satisfy%
\[
\frac{\partial\left|  \Psi(x,y,t)\right|  ^{2}}{\partial t}+\frac
{\partial\left(  \left|  \Psi(x,y,t)\right|  ^{2}\dot{x}\right)  }{\partial
x}+\frac{\partial\left(  \left|  \Psi(x,y,t)\right|  ^{2}\dot{y}\right)
}{\partial y}=0
\]
from which we deduce the (non-standard) guidance equations\footnote{For
standard Hamiltonians, $\dot{x}=j/\left|  \psi\right|  ^{2}$ usually reads
$\dot{x}=(\hbar/m)\operatorname*{Im}(\nabla\psi/\psi)$. Here the velocity
field is unusual because the Hamiltonian is.} $\dot{x}=0,\;\dot{y}=ax$ and the
de Broglie-Bohm trajectories $x(t)=x_{0},\;y(t)=y_{0}+ax_{0}t$.

Now the initial product wavefunction $\Psi_{0}(x,y)=\psi_{0}(x)g_{0}(y)$
evolves into the entangled wavefunction $\Psi(x,y,t)=\psi_{0}(x)g_{0}(y-axt)$.
In the limit $at\rightarrow0$, we have $\Psi(x,y,t)\approx\psi_{0}(x)g_{0}(y)$
and the system wavefunction $\psi_{0}(x)$ is undisturbed. Yet, no matter how
small $at$ may be, at the hidden-variable level the `pointer' position
$y(t)=y_{0}+ax_{0}t$ contains information about the value of $x_{0}$ (and of
$x(t)=x_{0}$). And this `subquantum' information about $x$ will be visible to
us if the initial pointer distribution $\pi_{0}(y)$ is sufficiently narrow.

For consider an ensemble of similar experiments, where $x$ and $y$ have the
initial joint distribution $P_{0}(x,y)=\left|  \psi_{0}(x)\right|  ^{2}\pi
_{0}(y)$ (equilibrium for $x$ and nonequilibrium for $y$). The continuity
equation%
\[
\frac{\partial P(x,y,t)}{\partial t}+ax\frac{\partial P(x,y,t)}{\partial y}=0
\]
implies that at later times $P(x,y,t)=\left|  \psi_{0}(x)\right|  ^{2}\pi
_{0}(y-axt)$. If $\pi_{0}(y)$ is localised -- say $\pi_{0}(y)\approx0$ for
$\left|  y\right|  >w/2$ -- then from a standard measurement of $y$ we may
deduce that $x$ lies in the interval $(y/at-w/2at,\;y/at+w/2at)$ (so that
$P(x,y,t)\neq0$), where the error margin $w/2at\rightarrow0$ as the width
$w\rightarrow0$.

Thus, if the nonequilibrium `apparatus' distribution $\pi_{0}(y)$ has an
arbitrarily small width $w$, then to arbitrary accuracy we may measure the
position $x$ of each equilibrium particle without disturbing the wavefunction
$\psi_{0}(x)$.\footnote{For finite $w<\Delta$, where $\Delta$ is the width of
$\left|  g_{0}(y)\right|  ^{2}$, we may make probabilistic statements about
the value of $x$ that convey more information than quantum theory allows;
while if $w>\Delta$, the measurements will be less accurate than those of
quantum theory [7].}

We have for simplicity considered an exactly-solvable system with an unusual
total Hamiltonian. Similar conclusions hold for more standard systems: if the
interaction between $x$ and $y$ is sufficiently weak, then while $\psi_{0}(x)$
is hardly disturbed, at the hidden-variable level $y$ generally contains
information about $x$ -- information that is visible if $y$ has a sufficiently
narrow distribution.

Generalising, if $w$ is arbitrarily small, then by a sequence of such
measurements, it is clear that for a system particle with arbitrary
wavefunction $\psi(x,t)$ we can determine the hidden trajectory $x(t)$ without
disturbing $\psi(x,t)$, to arbitrary accuracy.

\section{Distinguishing Non-Orthogonal Quantum States}

In quantum mechanics, non-orthogonal states $\left|  \psi_{1}\right\rangle $,
$\left|  \psi_{2}\right\rangle $ (with $\langle\psi_{1}|\psi_{2}\rangle\neq0$)
cannot be distinguished without disturbing them [19]. This theorem breaks down
in the presence of nonequilibrium matter [7].

For example, if $\left|  \psi_{1}\right\rangle $, $\left|  \psi_{2}%
\right\rangle $ are distinct initial states of a single spinless particle,
then in de Broglie-Bohm theory the velocity fields $j_{1}(x,t)/\left|
\psi_{1}(x,t)\right|  ^{2}$, $j_{2}(x,t)/\left|  \psi_{2}(x,t)\right|  ^{2}$
generated by the wavefunctions $\psi_{1}(x,t)$, $\psi_{2}(x,t)$ will in
general be different, even if $\langle\psi_{1}|\psi_{2}\rangle=\int
dx\,\,\psi_{1}^{\ast}(x,0)\psi_{2}(x,0)\neq0$. The hidden-variable
trajectories $x_{1}(t)$ and $x_{2}(t)$ -- associated with $\psi_{1}(x,t)$ and
$\psi_{2}(x,t)$ respectively -- will generally differ if $\psi_{1}%
(x,0)\neq\psi_{2}(x,0)$ (even if $x_{1}(0)=x_{2}(0)$). Thus, a subquantum
measurement of the particle trajectory (even over a short time) would
generally enable us to distinguish the quantum states $\left|  \psi
_{1}\right\rangle $ and $\left|  \psi_{2}\right\rangle $ without disturbing
them, to arbitrary accuracy.

\section{Eavesdropping on Quantum Key Distribution}

Alice and Bob want to share a secret sequence of bits that will be used as a
key for cryptography. During distribution of the key between them, they must
be able to detect any eavesdropping by Eve. Three protocols for quantum key
distribution -- BB84 [20], B92 [21], and E91 (or EPR) [22] -- are known to be
secure against classical or quantum attacks (that is, against eavesdropping
based on classical or quantum physics) [23]. But these protocols are
\textit{not} secure against a `subquantum' attack [7].

Both BB84 and B92 rely on the impossibility of distinguishing non-orthogonal
quantum states without disturbing them. In BB84 Alice sends Bob a random
sequence of spin-1/2 states $\left|  +z\right\rangle ,\;\left|
-z\right\rangle ,\;\left|  +x\right\rangle ,\;\left|  -x\right\rangle $, while
in B92 she sends a random sequence of arbitrary non-orthogonal states $\left|
u_{0}\right\rangle ,\;\left|  u_{1}\right\rangle $ (the states being subjected
to appropriate random measurements by Bob). In each case the sequence is
chosen by Alice. But if Eve possesses non-quantum matter with an arbitrarily
narrow nonequilibrium distribution, she may identify the states sent by Alice
without disturbing them, to arbitrary accuracy, and so read the supposedly
secret key. (For B92, $\left|  u_{0}\right\rangle ,$ $\left|  u_{1}%
\right\rangle $ could be states of a spinless particle with wavefunctions
$\psi_{0}(x,t),$ $\psi_{1}(x,t)$, which Eve may distinguish by monitoring the
hidden-variable trajectories. Similarly for BB84 -- though for spin-1/2 states
one must consider pilot-wave theory for two-component wavefunctions [7, 10].)

E91 is particularly interesting for it relies on the completeness of quantum
theory -- that is, on the assumption that there are no hidden `elements of
reality'. Pairs of spin-1/2 particles in the singlet state are shared by Alice
and Bob, who perform spin measurements along random axes. For coincident axes
the same bit sequence is generated at each wing, by apparently random quantum
outcomes. `The eavesdropper cannot elicit any information from the particles
while in transit ..... because there is no information encoded there' [22].
But our Eve has access to information outside the domain of quantum theory.
She can measure the particle positions while in transit, without disturbing
the wavefunction, and so \textit{predict} the outcomes of spin measurements at
the two wings (for the publicly-announced axes).\footnote{In Bell's pilot-wave
theory of spin-1/2 [10], particle positions within the wavepacket determine
the outcomes of Stern-Gerlach measurements.} Thus Eve is able to predict the
key shared by Alice and Bob.

\section{Outpacing Quantum Computation}

Quantum theory allows parallel Turing-type computations to occur in different
branches of the state vector for a single computer [24]. However, owing to the
effective collapse that occurs under measurement, an experimenter is able to
access only one result; the outputs of the other computations are lost. Of
course, by clever use of entanglement and interference, one can make quantum
computation remarkably efficient for certain special problems. But in general,
what at first sight seems to be a massive increase in computational power is
not, in fact, realised in practice.

All the results of a parallel quantum computation could be read, however, if
we had access to nonequilibrium matter with a very narrow distribution [3, 7].

For each result could be encoded in an integer $n$, and stored as an energy
eigenvalue $E_{n}$ for a single spinless particle (a component of the
computer). At the end of the computation the particle wavefunction will be a
superposition%
\[
\psi(x,t)=\sum_{n\in S}\phi_{n}(x)e^{-iE_{n}t}%
\]
of $N$ energy eigenfunctions $\phi_{n}(x)$, where $S$ is an \textit{unknown}
set of $N$ quantum numbers. (We assume a Hamiltonian $\hat{H}=\hat{p}%
^{2}/2+V(\hat{x})$, where the mass $m=1$ and $\hbar=1$.) In standard quantum
theory an energy measurement for the particle yields just one value $E_{n}$.
To find out what other eigenvalues are present, one would have to run the
whole computation many times -- to produce an ensemble of copies of the same
wavefunction -- and repeat the energy measurement for each. And so one may as
well just run many computations on a single classical computer, one after the other.

But the hidden-variable particle trajectory $x(t)$ -- determined by $\dot
{x}(t)=j/\left|  \psi\right|  ^{2}$ or $\dot{x}=\operatorname*{Im}\left(
\nabla\psi/\psi\right)  $ -- contains information about all the modes in the
superposition (provided the $\phi_{n}(x)$ overlap in space). If we had a
sample of nonequilibrium matter with a very narrow distribution, we could use
it to measure $x(t)$ without disturbing $\psi(x,t)$. We could then read the
set $S$ of quantum numbers: having measured the values of $x(t),\;\dot{x}(t)$
at $N$ times $t=t_{1},\;t_{2},\;....,\;t_{N}$, the equation $\dot
{x}=\operatorname*{Im}\left(  \nabla\psi/\psi\right)  $ may be solved for the
$N$ quantum numbers $n$.\footnote{We are assuming that $\phi_{n}(x),\;E_{n}$
are known functions of $x,\;n$ -- obtained by solving the eigenvalue problem
$\hat{H}\phi_{n}(x)=E_{n}\phi_{n}(x)$. The $N$ pairs of values $x(t_{i}%
),\;\dot{x}(t_{i})$ might be obtained by subquantum measurements of $x(t)$ at
$2N$ times $t=t_{1},\;t_{1}+\epsilon,\;t_{2},\;t_{2}+\epsilon,\;....,\;t_{N},$
$t_{N}+\epsilon$, with $\epsilon$ very small.} Thus we could read the results
of all $N$ arbitrarily long parallel computations (at the price of solving $N$
simultaneous equations), even though the computer has been run only once.

By combining subquantum measurements with quantum algorithms, we could solve
$NP$-complete problems in polynomial time, and so outpace all known quantum
(or classical) algorithms [7].

To see this, consider the computational enhancement noted by Abrams and Lloyd
in nonlinear quantum mechanics [25]. Let a quantum (equilibrium) computer
begin with $n+1$ qubits in the state $\left|  00\;.....\;0\right\rangle $ and
apply the Hadamard gate $H$ (which maps $\left|  0\right\rangle \rightarrow
(\left|  0\right\rangle +\left|  1\right\rangle )/\sqrt{2}$, $\left|
1\right\rangle \rightarrow(\left|  0\right\rangle -\left|  1\right\rangle
)/\sqrt{2}$) to each of the first $n$ qubits to produce $(1/\sqrt{2^{n}}%
)\sum_{x}\left|  x,\;0\right\rangle $, where the $n$-bit `input' $x$ ranges
from $00\;.....\;0$ to $11\;.....\;1$ (or from $0$ to $2^{n}-1$). Then use an
`oracle' or `black box' to calculate -- in parallel -- a function $f(x)=0$ or
$1$, whose value is stored in the last qubit, producing $(1/\sqrt{2^{n}}%
)\sum_{x}\left|  x,\;f(x)\right\rangle $. Applying $H$ again to each of the
first $n$ qubits produces a state containing the term $(1/2^{n})\sum
_{x}\left|  00\;.....\;0,\;f(x)\right\rangle $. If upon quantum measurement of
the first $n$ qubits we obtain $\left|  00\;.....\;0\right\rangle $, the total
effective state becomes $\left|  00\;.....\;0\right\rangle \otimes\left|
\psi\right\rangle $ where $\left|  \psi\right\rangle \propto\left|
0\right\rangle (2^{n}-s)/2^{n}+\left|  1\right\rangle s/2^{n}$ and $s$ is the
number of inputs $x$ such that $f(x)=1$ (the total number of inputs being
$2^{n}$).\footnote{The quantum equilibrium probability of obtaining $\left|
00\;.....\;0\right\rangle $ is at least 1/4 [25].} As Abrams and Lloyd point
out, we could solve $NP$-complete problems if we could distinguish between
$s=0$ and $s>0$ for the state $\left|  \psi\right\rangle $ of the last qubit.
This could be accomplished by nonlinear evolution, in which non-orthogonal
states evolve into (distinguishable) orthogonal ones [25]. But equally,
non-orthogonal qubits could be distinguished using our nonequilibrium matter.
Here, the de Broglie-Bohm trajectory $x(t)$ of an equilibrium particle guided
by $\psi(x,t)=\langle x|\psi(t)\rangle$ will in general be sensitive to the
value of $s$, which may therefore be read by a subquantum measurement of
$x(t)$ [7].

\section{Conclusion}

We have argued that immense physical resources are hidden from us by quantum
noise, and that we will be unable to access those resources only for as long
as we are trapped in the `quantum heat death' -- a state in which all systems
are subject to the noise associated with the Born probability distribution
$\rho=|\psi|^{2}$.

It is clear that hidden-variables theories offer a radically different
perspective on quantum information theory. In such theories, a huge amount of
`subquantum information' is hidden from us simply because we happen to live in
a time and place where the hidden variables have a certain `equilibrium'
distribution. As we have mentioned, nonequilibrium instantaneous signals occur
not only in pilot-wave theory but in \textit{any} deterministic
hidden-variables theory [15, 16]. And in pilot-wave theory at least, we have
shown that the security of quantum cryptography depends on our being trapped
in quantum equilibrium; and, that nonequilibrium would unleash computational
resources far more powerful than those of quantum computers.

Some might prefer to regard this work as showing how the principles of quantum
information theory depend on a particular axiom of quantum theory -- the Born
rule $\rho=|\psi|^{2}$. (One might also consider the role of the axiom of
linear evolution [25, 26].)

But if one takes hidden-variables theories seriously as physical theories of
Nature, one can hardly escape the conclusion that we just happen to be
confined to a particular state in which our powers are limited by an
all-pervading statistical noise. It then seems important to search for
violations $\rho\neq|\psi|^{2}$ of the Born rule [3--7].

\textbf{Acknowledgement.} This work was supported by the Jesse Phillips Foundation.

\begin{center}
\textbf{REFERENCES}
\end{center}

[1] A. Valentini, Phys. Lett. A \textbf{156}, 5 (1991).

[2] A. Valentini, Phys. Lett. A \textbf{158}, 1 (1991).

[3] A. Valentini, PhD thesis, International School for Advanced Studies,
Trieste, Italy (1992).

[4] A. Valentini, in \textit{Bohmian Mechanics and Quantum Theory: an
Appraisal}, eds. J. T. Cushing \textit{et al.} (Kluwer, Dordrecht, 1996).

[5] A. Valentini, in \textit{Chance in Physics: Foundations and Perspectives},
eds. J. Bricmont \textit{et al}. (Springer, Berlin, 2001) [quant-ph/0104067].

[6] A. Valentini, Int. J. Mod. Phys. A (forthcoming).

[7] A. Valentini, \textit{Pilot-Wave Theory of Physics and Cosmology}
(Cambridge University Press, Cambridge, forthcoming).

[8] L. de Broglie, in \textit{\'{E}lectrons et Photons: Rapports et
Discussions du Cinqui\`{e}me Conseil de Physique}, ed. J. Bordet
(Gauthier-Villars, Paris, 1928). [English translation: G. Bacciagaluppi and A.
Valentini, \textit{Electrons and Photons: The Proceedings of the Fifth Solvay
Congress} (Cambridge University Press, Cambridge, forthcoming).]

[9] D. Bohm, Phys. Rev. \textbf{85}, 166; 180 (1952).

[10] J. S. Bell, \textit{Speakable and Unspeakable in Quantum Mechanics}
(Cambridge University Press, Cambridge, 1987).

[11] P. Holland, \textit{The Quantum Theory of Motion: an Account of the de
Broglie-Bohm Causal Interpretation of Quantum Mechanics} (Cambridge University
Press, Cambridge, 1993).

[12] D. Bohm and B. J. Hiley, \textit{The Undivided Universe: an Ontological
Interpretation of Quantum Theory} (Routledge, London, 1993).

[13] J. T. Cushing, \textit{Quantum Mechanics: Historical Contingency and the
Copenhagen Hegemony} (University of Chicago Press, Chicago, 1994).

[14] \textit{Bohmian Mechanics and Quantum Theory: an Appraisal}, eds. J. T.
Cushing \textit{et al.} (Kluwer, Dordrecht, 1996).

[15] A. Valentini, Phys. Lett. A (in press) [quant-ph/0106098].

[16] A. Valentini, in \textit{Modality, Probability, and Bell's Theorems},
eds. T. Placek and J. Butterfield (Kluwer, Dordrecht, 2002) [quant-ph/0112151].

[17] D. Bohm and B. J. Hiley, Found. Phys. \textbf{14}, 255 (1984).

[18] A. Valentini, Phys. Lett. A \textbf{228}, 215 (1997).

[19] M. A. Nielsen and I. L. Chuang, \textit{Quantum Computation and Quantum
Information} (Cambridge University Press, Cambridge, 2000).

[20] C. H. Bennett and G. Brassard, in \textit{Proceedings of IEEE
International Conference on Computers, Systems and Signal Processing,
Bangalore, India} (IEEE, New York, 1984).

[21] C. H. Bennett, Phys. Rev. Lett. \textbf{68}, 3121 (1992).

[22] A. Ekert, Phys. Rev. Lett. \textbf{67}, 661 (1991).

[23] N. Gisin \textit{et al}., Rev. Mod. Phys. (forthcoming) [quant-ph/0101098].

[24] D. Deutsch, Proc. Roy. Soc. London A \textbf{400}, 975 (1985).

[25] D. S. Abrams and S. Lloyd, Phys. Rev. Lett. \textbf{81}, 3992 (1998).

[26] A. Valentini, Phys. Rev. A \textbf{42}, 639 (1990).
\end{document}